# Stablecoins and Central Bank Digital Currencies: Policy and Regulatory Challenges
Barry Eichengreen and Ganesh Viswanath-Natraj[1]
November 2021

## 1. Introduction

Digital currencies are coming, not least (or, some would say, especially) in Asia. The question is how quickly and in what form.

Digital coins and currencies come in three basic flavors: plain-vanilla cryptocurrencies such as Bitcoin; stablecoins such as Tether and USD Coin; and central bank digital currencies (CBDCs). In what follows we contribute to the literatures on stablecoins and CBDCs. It makes sense to analyze these alternative units together. Stablecoin backers are seeking to appropriate and privatize the payments function traditionally dominated by central bank money. In turn, central banks are contemplating issuance of CBDCs, in part, precisely in order to better compete with stablecoins and prevent this privatization of payments. This dynamic leads immediately to a set of interrelated questions. Do stablecoins have significant advantages over existing payments mechanisms? If they are less than stable, do pose risks to the integrity of the payments system? Are central banks right to resist privatization of this function? And, if they are right, is CBDC issuance the best way of doing so?

A starting point for analyzing stablecoins is to observe that they come in different "flavors." Some stablecoins are fully backed by legal-tender money to which they are pegged, while others are only partially or fractionally backed by legal-tender money or equivalent liquid assets.[2] A stablecoin fully backed by cash and short-term Treasuries should be immune from run risk so long as Treasuries retain their safe-asset status.[3] Not so a stablecoin backed in part by commercial paper and other typically liquid instruments that can become illiquid under strained circumstances.

In this paper, we provide time-varying estimates of that risk for the leading stablecoin, Tether, using data from the Tether/USD futures market. We show that devaluation risk is priced, though it is relatively small. On the assumption that in the event that the peg is broken investors in Tether lose the full value of investments, we show that the average probability of this event is priced on average at 0.3 percent in annualized terms.

---

[1] University of California, Berkeley and University of Warwick, respectively. This draft was prepared for the Asian Economic Panel. It draws on earlier publications (Eichengreen 2021a,b).
[2] Our focus is on centralized stablecoins backed by dollar assets, such as Tether and USD Coin. In addition there are "cryptocurrency collateralized" stablecoins such as MakerDAO's DAI token. DAI tokens are generated when an investor deposits a set amount of collateral (typically Ethereum) into a collateralized debt position (CDP). While this stablecoin avoids the risk of a centralized custodian of assets, risky collateral can lead to instability of the peg (Kozhan and Viswanath-Natraj, 2021). A second type, "algorithmic stablecoins," hold no backing but rely on an algorithm to maintain their stability against the dollar. Tether, the leading stablecoin, holds half its reserves in commercial paper. In August 2021 USD Coin moved from holding just 60 percent of its reserves in dollar bank accounts, and the rest in sundry (unspecified) bonds and debt instruments to holding 100 percent in cash and short-duration U.S. Treasury bonds. In May 2021 Tether published a report showing that 2.9 percent of its collateral or backing was held in cash, while 49.5 percent was held in commercial paper, 12.6 percent in secured loans, and 10 percent in corporate bonds, funds and precious metals.
[3] That safe-asset status is not guaranteed, of course. Group of Thirty (2021) warns that more episodes of market disfunction like that which occurred in March 2020 could damage that haven status.

When analyzing the correlates of default or run risk, we find that its time variation fluctuates mainly with the intra-day volatility of Bitcoin prices. For example, the peak annualized default probability of 2.0 percent occurred on "Black Thursday," March 12$^{th}$, 2020, when major cryptocurrencies such as Bitcoin fell by some 50 percent. This pattern suggests that stablecoins are not used as a generalized means of payment – the function to which stablecoin purveyors aspire – but rather as a specialized gateway for buying and selling Bitcoin and its substitutes.

We turn next to central bank digital currencies. After reviewing the cases for and against, we focus specifically on implications for the international monetary system. We distinguish two questions. First, is digitization an advantage in the competition for dominant international and reserve currency status? Specifically, will the fact that China is moving faster than the United States toward issuing a CBDC tilt the international monetary playing field away from the dollar and toward the renminbi? Second, are CBDCs subversive to the very notion of dominant international and reserve currencies? Will they allow countries to transact directly amongst themselves, using their own national currencies, without having to go through the dollar as at present? In this brave new digital world, will all currencies share the dollar's "exorbitant privilege?"

Our answers, in brief, are no and no. Moving from a renminbi held in the form of cash, bank balances and short-term securities to a digital renminbi held in a smartphone wallet or retail account with the People's Bank of China will not, by itself, make the currency significantly more attractive for use in cross-border transactions. It will not remove other obstacles – China's retention of capital controls and questions about anonymity and surveillance – that have slowed the pace of renminbi internationalization. To avoid undermining the operation of controls, renminbi held and transferred between digital wallets will be limited to small balances. And claims of limited surveillance – that the PBOC will track and gather only limited information about transactions completed using its digital currency – are unlikely to be regarded as credible.

For the CBDCs of smaller countries to be usable in cross-border transactions, they will have to be made interoperable. While central banks and international organizations are exploring this possibility, interoperability faces formidable obstacles. Widespread interoperability would require either a very large number of bilateral agreements between central banks, each supported by the relevant financial infrastructure, or a multilateral organization akin to the International Monetary Fund or the World Trade Organization with powers to establish and operate the same. This, clearly, isn't going to happen.

2. **Stablecoin Basics**

Although stablecoins first came to prominence with Facebook's announcement in June 2019 of plans to launch a currency-basket based unit known as Facebook Coin, then Libra and most recently Diem, such instruments have in fact been around for much longer. The first such unit, Tether (originally Realcoin), with a one-to-one link to the dollar, was launched in 2014 and began trading in 2015.[4] The tokens themselves circulate on decentralized networks but are issued and redeemed by a Hong Kong-based entity Tether Ltd. Tokens can be purchased with actual dollars transferred from a bank account, debit card or the like on digital exchanges such as Bitfinex, Binance and Coinbase.

By 2021 the stablecoin universe had become crowded. Table 1 lists the ten leading stablecoins by market capitalization. Tether is a particularly interesting case because it has been around the longest, because its market capitalization is the highest, and because its colorful history illustrates some of the controversial issues surrounding stablecoins. Although Tether has long claimed to hold a dollar bank deposit as backing for each dollar token issued, this does not appear to have continuously been the case, for the simple reason that for portions of its history Tether has not had a banking partner.[5] The Attorney General of the State of New York alleged that Tether misrepresented the extent of its backing and, together with Bitfinex, covered up the loss of $850 million in customer funds. The suit was settled in 2021, when Tether and Bitfinex agreed to pay fines of $18.5 million and cease all trading with New York-based residents and entities. As part of the settlement, Tether is required to provide quarterly financial reports to the State Attorney General. It is required to document that it is properly segregating the accounts of its clients and company executives and specifically to report all transfers between Bitfinex and Tether.[6]

Tether's balance sheet looks a lot like that of Primary Reserve Fund, the commercial paper-holding money market mutual fund forced to "break the buck" (break its one-to-one peg to the dollar) in 2008. The risk, as was learned during the Global Financial Crisis, is that liquid commercial paper can abruptly become illiquid. As Table 1 shows, the market capitalization of the four largest US dollar stablecoins, at more than $1 trillion, already approaches that of the largest institutional money fund, JPMorgan Prime Money Market Fund. Were a panic to force these stablecoin issuers to abruptly sell off a significant share of their commercial paper and corporate bond holdings, the liquidity of those markets could be impaired.[7] Officials' concern with the stability of the commercial paper market thus provides another basis for prospective regulation.

How likely is a stablecoin such as Tether, USD Coin or Diem to transform the payments landscape? Technically, there is no obstacle to customers downloading Facebook's digital wallet and using it to transact in the Paxos Dollar or, prospectively, Diem, Facebook's dollar-linked stablecoin.[8] The technology exists.[9] But the regulatory approval doesn't. We have already seen multiple potential reasons why regulators may hesitate to provide that approval, from threats to the stability of the commercial paper market to difficulty of enforcing know-your-customer rules designed to prevent money laundering and other elicit transactions. Take-up may also be slowed if there are doubts about the continued stability of the stablecoin. It is to this question that we now turn.

---

[4] By 2021 there were four Tether stablecoins: U.S. dollar tied, euro tied, offshore yuan tied and gold tied; although trading was and is dominated by the dollar-tied token.

[5] Initially, it held accounts with banks in Taiwan and used Wells Fargo as correspondent to transfer funds to and from, inter alia, U.S. investors. Starting in March 2017, however, Wells Fargo refused to process additional U.S. dollar wire transfers. From this point Tether apparently held incoming funds in an account in its general counsel's name at the Bank of Montreal. Later it held funds in a comingled bank account in Puerto Rico maintained by Bitfinex, the main exchange on which Tether is traded. (Tether's accounting referred to these deposits as "receivables" owed it by Bitfinex.) An audit of these accounts, to be undertaken by Friedman LLP, a New York-based accounting, tax and business consulting firm, in 2017, was never completed.

[6] Tether and Bitfinex neither admitted nor denied the Attorney General's findings.

[7] A speculative attack on Tether occurred on 11th October, 2018 on the exchange Bitfinex, when Tether transaction prices fell to 95 cents. On that day, Bitfinex decided to temporarily pause national-currency deposits (USD, GBP, EUR, JPY) for certain customer accounts in the face of processing complications. This triggered peg discounts as investors speculated that the peg was under-collateralized or otherwise could not meet redemptions.

### 3. Stablecoin Default Risk

In this section we derive an estimate of Tether default risk. We use data on Tether/USD (USDT) futures, which have traded on the FTX derivatives exchange since February 2020. Data on USDT spot and futures prices from https://www.coinapi.io/.[10] Summary statistics for these and related data are shown in Tables 2 and 3. Note that Tether futures typically trade at a discount relative to spot, as if investors perceive a risk of default. We define $s_t$ and $f_t$ as the spot and futures price of a unit of Tether, and $\Delta_t$ as the peg-price premium $s_t$ - 1.

Assume that the price dynamics of the peg follow an AR(1) process with a mean reversion coefficient $\rho$ as in eq. 1. (Stability requires $\rho < 1$.[11])

$$\Delta_{t+1} = \rho \Delta_t + \epsilon_{t+1}, 0 < \rho < 1 \qquad (1)$$

Iterating the expression in equation (1) forward, we obtain an expression for the peg-price deviation at expiry of the contract (at time $t + h$). This is the current deviation discounted by the mean reversion coefficient $\rho$, plus a discounted sum of Tether-specific shocks $\epsilon_{t+s}$.

$$\Delta_{t+h} = \rho^h \Delta_t + \sum_{s=1}^{h} \rho^{h-s} \epsilon_{t+s} \qquad (2)$$

With probability $P$, the peg to the dollar is broken. For simplicity, assume for the moment that investors recover 0 percent of their funds. With probability $1 - P$, the spot rate is equal to an exponential decay of peg-price deviations, including shocks that are discounted by the mean reversion coefficient $\rho$. The spot rate at expiry is given by:

$$s_{t+h} = \begin{cases} 1 + \rho^h \Delta_t + \sum_{s=1}^{h} \rho^{h-s} \epsilon_{t+s}, & \text{with probability } 1 - P \\ 0, & \text{with probability } P \end{cases} \qquad (3)$$

---

[8] Novi was launched in October 2021 for use with the Paxos Dollar, with the crypto exchange Coinbase providing custody services.

[9] Additional issues arise when seeking to use these units outside the jurisdiction in whose currency they are denominated. Money laundering and related concerns become even more of an issue in the context of cross-border transactions. There is the danger that a country's capital controls will be more easily evaded or that it may become subject to large-scale currency substitution. Residents of countries where the U.S. dollar is not widely used will want to convert their dollar-linked stablecoins into local currency, requiring the trading services of commercial banks or automated dealers, raising additional regulatory issues. We discuss these questions at more length in the penultimate section of the paper on central bank digital currencies and the international monetary system, since many of the same issues arise there.

[10] Coinapi offers a monthly subscription that provides access to their api, which includes historical cryptocurrency data. Price data are available from the FTX exchange.

[11] The coefficient $\rho$ can also be used to estimate the half-life of the system.

The expectations hypothesis states that the futures price for a contract expiring $h$ periods from now equals the expectation of the spot rate $h$ periods from now.

$$f_t = E_t[S_{t+h}] \qquad (4)$$

Iterating forward $h$ periods, we can show that, conditional on the peg being in the "no default" state, peg-price deviations dissipate with time: $\rho^h$ goes to zero as the expiry time of the contract $h$ goes to infinity.

$$f_t = (1 - \mathcal{P}) \times E_t[S_{t+h}|\text{No Default}] + \mathcal{P} \times E_t[S_{t+h}|\text{Default}] \qquad (5)$$
$$= (1 - \mathcal{P}) \times (1 + \rho^h \Delta_t)$$

Substituting $s_t = 1 + \Delta_t$ we obtain an expression for the probability of default risk:

$$\mathcal{P}_t = 1 - \frac{f_t}{1 + \rho^h(s_t - 1)} \qquad (6)$$

The default probability is increasing in the spot rate and decreasing in the futures rate. It is inversely related to the futures-spot basis $f_t - s_t$.[12] This approach allows us to impute a time-varying probability of default from time variation in the futures-spot basis. Time series of the futures-spot basis is seen in Figure 1. The spikes toward the left correspond to March 12th, 2020, when Bitcoin prices dropped by 50 percent.

To compute the default probability, we set the autoregressive parameter $\rho$ in eq. 1 to 0.73, its average over the sample. In calculating the annualized probability, we set the horizon of the futures contract $h = 90$. Subject to these parameters, Table 3 also shows the implied probability of default, which averages 0.3 percent over the period. Figure 2 shows its time series.[13]

Table 4 analyzes the determinants of this default probability, which is regressed against the intra-day volatility of Bitcoin and USDT, as well as daily returns of Bitcoin. Both Bitcoin volatility and USDT volatility are positively associated with default risk. Whereas a 100 basis point increase in Bitcoin volatility increases the probability of default by 4.3 basis points, a 100 basis point increase in USDT volatility increases it by 21.2 basis points. When we include both variables only Bitcoin volatility is significant.

Why is Bitcoin volatility important? Bitcoin is the leading cryptocurrency by market capitalization, depending on currently prevailing prices, in the neighborhood of $650 billion.[14] Bitcoin's ups and downs are highly visible to investors, so that when holding Bitcoin becomes less attractive other digital currencies are similarly perceived as less attractive. A more important explanation may be that a significant share of the demand for Tether is as a vehicle currency in the crypto market. Selling Bitcoin on an exchange and having the proceeds transferred to one's bank can be complicated and time-consuming; exchanging Bitcoin for Tether is quicker and easier. This helps to explain why there is a demand for Tether and other stablecoins during risk-

---

[12] As the horizon of the futures contract approaches infinity, the equation simplifies to $P = 1 - f_t$. (the probability is equal to the futures discount relative to the peg).
[13] We have trimmed the series lower bound to zero, since negative default probabilities (of which there are few) make no sense in the context of our model.
[14] Tether's market cap, recall, is roughly one-tenth this amount.

off events.[15] Stablecoins are mainly used as a vehicle for getting in and out of crypto markets. Our essential point is that investors are systematically pricing a significant risk of Tether default and that this risk is a function of Bitcoin volatility and systemic risk in the cryptocurrency market.[16]

Why aren't these imputed default probabilities higher? One answer is that significant redemption costs limit the risk of a run in which investors seek to redeem Tether at a rate faster than it can liquidate its commercial paper holdings at something approaching par value. Whereas Tether charges a fee of 0.1 percent for fiat deposits, the fee attached to fiat withdrawal is 0.1 percent of the amount withdrawn, *with a minimum of $1,000* (Hampl and Gyonyorova 2020).[17] This is directly analogous to the rules for money market mutual funds adopted by the U.S. Securities and Exchange Commission in response to the Primary Reserve Fund affair in 2008, giving funds the discretion to introduce fees of up to 2 percent on redemptions (or restrict redemptions for up to 10 business days) in the event that a fund is unable to raise sufficient liquidity to meet shareholder redemptions.[18]

Looking forward, public officials have raised questions about stablecoins and their regulation. Insofar stablecoins are simply the digital equivalent of prime money market funds, which similarly invest in high-quality commercial paper, they raise the same financial stability issues.[19] In cases where they have only partial reserve backing, they raise the same issues as fractional reserve banks. As Federal Reserve Chair Jerome Powell put it in July 2021, "If we're going to have something that looks just like a money-market fund, or a bank deposit, a narrow bank, and it's growing really fast, we really out to have appropriate regulation – and today we don't."[20]

In 2020 the European Commission published a proposal for regulation of crypto assets, including stablecoins, which it refers to as E-Money Tokens. The most important provision would require issuers to submit whitepapers describing their proposed token in advance to national financial supervisors and empower the latter to prohibit or authorize issuance. In the U.S., the "Stablecoin Tethering and Bank Licensing Enforcement Act" introduced as a bill by three members of the U.S. Congress in December 2020 would require any prospective stablecoin issuer to obtain a banking charter, and thus be subject to all the same capital, liquidity, disclosure and other requirements, not to mention oversight, as banks.[21]

---

[15] Lyons and Viswanath-Natraj (2020) find evidence that stablecoins exhibit safe have properties during a crypto risk-off event, in which investors liquidate Bitcoin into stablecoins as a store of value.

[16] Less weight should perhaps be attached to our numerical estimates of perceived default risk, which may be too high or too low for a number of reasons. Most obviously, translating our calculations based on spot/forward differentials into perceived default probabilities requires an assumption about post-default recovery rates, set here for illustrative purposes at 0 percent. Were the recovery rate instead 75 percent, the average imputed default risk would be 1.2 percent (not 0.3 percent, as above). Were the recovery rate 90 percent, average default risk in annualized terms would instead be imputed as 3 percent.

[17] Routledge and Zetlin-Jones (2021) propose a similar rule under which conversion of the stablecoin into legal-tender cash would be temporarily suspended or honored only at a depreciated rate in response to traders' conversion demands. Temporarily suspending conversion of the stablecoin into cash would remove the incentive to run by eliminating the sequential service constraint, in the same way that temporarily suspending the convertibility of bank deposits into currency removes the danger of a bank run (Diamond and Dybvig 1983).

[18] The SEC first issued rules in 2010 requiring money market funds to maintain 10 percent of their portfolios in securities maturing daily and 30 percent in securities maturing within a week. In 2014 it then allowed these fees and gates to be imposed when weekly liquidity fell below 30 percent and requiring such fees when it fell below 10 percent.

[19] The quality of the commercial paper held by stablecoins is uncertain; Tether for one provides no details.

[20] Powell testimony is available at https://www.c-span.org/video/?513254-1/federal-reserve-chair-testifies-monetary-policy.

4. **CBDC Basics**

The technical and economic issues around CBDCs are well known. A central bank could issue a digital token that is loaded and stored on a wallet or app on a user's smartphone. Those tokens could then be used to complete transactions with other users, on a platform or network operated by the central bank or its designee. Or the central bank could issue a token-less digital currency by opening retail accounts for individuals and nonfinancial firms, making digital deposits there, and executing transactions on instruction. Those transactions would again utilize a platform or network operated by the central bank, conceivably the already-existing real-time gross settlement system through which commercial bank transfers are cleared.

Yet a third option would be for the CBDC to be distributed by other intermediaries. The central bank would distribute CBDC to, say, commercial banks that would open digital currency accounts for their customers, much as they currently distribute paper currency notes via automatic teller machines. One can imagine both indirect systems in which the CBDC would be a liability of commercial banks, backed by their reserve accounts at the central bank, and hybrid systems in which the CBDC would be a liability of the central bank but commercial banks would be customer facing.

Digital transactions could be secured and verified using blockchain, conceivably a public blockchain where anyone is allowed to contribute to the verification, but more plausibly either a private blockchain where only authorized entities are allowed to participate and the operator can override or delete entries on the blockchain, or a permissioned blockchain where anyone is allowed to join after verification of their identity but different parties are permitted to perform only certain activities on the network.[22] Alternatively, central banks could opt to build the CBDC on top of the existing real-time gross settlement system and utilize a form of cryptography that doesn't involve blockchain at all.

Central banks will worry about the public-blockchain option, where anonymity and free entry may facilitate money laundering, tax evasion and terrorist finance. They will prefer the private and permissioned options, where the central bank itself or the central bank together with chartered commercial banks serve as nodes for verifying transactions. Others will worry about the private and permissioned options, which may expose private transactions to official scrutiny and, conceivably, interference. It also is unclear whether there are benefits of moving to distributed-ledger technology when transactions are still verified by a centralized authority, other than potential interoperability (i.e. compatibility) with other activities in the open blockchain sector (EU Blockchain Observatory and Forum p.35).

Central banks need to think about costs and benefits before moving in these directions. Most obviously, there is the straightforward financial cost of investing in new technology. There is the cost of investing in security and minimizing the danger that the payment system will be disrupted by hackers and digital terrorists. So long as there still exists paper currency, people will have means of purchasing essentials, such as food and water, even when electronic systems go down. In the absence of paper currency, this could become very much more difficult.[23] A CBDC would be a rich target. And if Mt. Gox can be hacked, why not a central bank?

---

[21] In addition, stablecoin issuers would be required to hold their reserves with the Federal Reserve System or else obtain insurance from the Federal Deposit Insurance Corporation.
[22] EU Blockhain Observatory and Forum (2021, p.52) similarly conclude that central banks are likely to prefer private or permissioned blockchains.

Finally, there are costs to the commercial banking system. If the interest rate paid on a retail account at the central bank is similar to that at commercial banks (which for the moment means that they need pay no interest at all), individuals may prefer the former for their greater safety, causing commercial banks to be disintermediated. The central bank will then find itself involved in private credit allocation. It will either have to go into the business of private lending itself, or else it will have to devise criteria for allocating the proceeds of its retail deposits to private financial institutions. There may also be heightened bank-run risk if depositors are free to flee to deposits at the central bank at the first hint of trouble. These risks could be addressed by limiting the amount of CBDC permitted for retail accounts to small denominations, but this in turn would limit the utility of the unit.

The potential benefits of CBDCs fall under two headings: transactions costs and financial inclusion. Completing a transaction electronically using a debit or credit card or online via one's bank requires possession of a bank account. 14 million American adults don't have one. Maintaining a bank account means paying the associated fees, since banks don't provide their services for free. Credit and debit-card transactions entail processing fees of up to 2 percent.[24] When using nonbank payments providers such as PayPal, it is similarly necessary to possess a bank account from which money can be transferred, or to possess a PayPal Cash card onto which cash can be loaded at a retailer, and for which the retailer will charge a fee. Funds transferred from PayPal to one's bank incur a fee of 1 percent of the transfer amount, while using PayPal to send funds to friends or family members using a debit or credit card incurs a fee of 2.9 percent. In developing countries, telecom operated payments system utilizing smartphones, such as M-Pesa, provide similar services. M-Pesa fees for fund transfer sim card users who have not registered with the service (registration requiring identify verification) approach 1 percent on low-value transactions.

A CBDC that was costless to use would relieve individuals of these 1 to 3 percent costs of electronic transactions. The central bank and government would provide basic electronic payments as a public service. It would treat them as a public utility. Whether 1-3 percent is big or small is in the eye of the beholder. Society would still be paying for the service, but the cost would be borne by the central bank out of seigniorage profits that it would otherwise presumably transfer to the Treasury, rather than by the end users. Total cost would only be less if one thinks that central banks can provide electronic payments services more efficiently than Visa or PayPal. If the problem is that credit-card and telecom companies are natural monopolies able to charge exorbitant fees for their services, then the direct way of solving this is to regulate those providers, not to create a central-bank supported competitor. In the U.S., Dodd-Frank already regulates debt-card fees, as noted, and M-Pesa is regulated by the financial services regulator (usually the central bank) of each country in which it operates.

---

[23] This begs the question of whether CBDC would entirely replace physical currency or merely supplement it.
[24] Debt-card fees are less and in the U.S. are regulated by the Dodd-Frank Wall Street Reform and Consumer Protection Act.

Arguments based on financial inclusion, though made by the likes of U.S. Treasury Secretary Janet Yellen, are weak. In the United States, the 6 percent of households without bank accounts and therefore also without debit or credit cards may lack sufficient income. But in order to operate the central bank's digital-currency app, they will have to have income sufficient to purchase a cellphone and a mobile-phone contract. Other unbanked individuals, lacking citizenship or engaged in legally dubious business practices, may prefer to fly under the radar, in which case they will also be reluctant to do business on the central bank's private or permissioned network.

Traditionally, the argument for public-policy initiatives to foster financial inclusion has been stronger in the developing-country context, where many individuals have lacked access to bank branches, automatic teller machines and so forth. But with the advent of cellphone-based services, as just described, this argument has lost its force. In addition, cellphone-based services such as M-Pesa are now using data gleaned from payments traffic to assign users credit scores and engage in micro-lending. It seems unlikely that central banks will want to (or should) get into the business of micro-lending to households. Transferring the electronic payments function from a commercial cellphone-based service to the central bank will presumably eliminate the capacity of the former to engage in micro-lending. In effect, this would be a step in the direction of financial exclusion.

5. **CBDCs and the International Monetary and Financial System**

The case for CBDCs would be stronger if these successfully brought down the cost of cross-border payments. International wire transfers generally incur fees of $50 or more. International ACH (automated clearinghouse) transfers have lower costs but can take three or more days to clear. For a cash transfer from storefront to storefront, the preferred vehicle of the unbanked, Western Union charges 7 percent for $100.

A central bank digital currency that was used globally could effect cross-border transactions more conveniently (no need to visit the Western Union store), more quickly, and at lower cost. A digital dollar that also circulated outside the United States, for example, or a Chinese CBDC that also circulated outside China would have this merit. If American importers as well as Chinese producers could obtain digital renminbi wallets, payment for orders could be seamlessly transferred from purchaser to supplier without mediation by correspondent banks or a clearinghouse.

Note, however, that fees for international payments are much lower, as a share of the funds transferred, for larger-value transactions. And other entities are already experimenting with digital technologies with the potential to reduce costs and accelerate transactions. Global banks such as Santander are using Ripple's open-source, semi-permissioned system to transfer funds between branches in different countries. SWIFT (the Society for Worldwide Interbank Financial Communication), through which most international interbank transfers are effected, is experimenting with distributed ledger technology. It has launched "Swift gpi," a set of high-speed electronic rails to increase the speed and predictability of high-value payments, and "SWIFT Go" for small payments. These systems allow participating banks (currently limited in number) to pre-validate information about the beneficiary, thereby avoiding costly and time-consuming mistakes, using an Application Programming Interface, or API, that allows the sending bank to automatically tap into information on the account of the receiving bank.

Similarly, countries with instant payment systems that do not use distributed-ledger technology but allow retail customers to transfer funds instantly between participating banks are exploring linking these up across countries. Singapore and Thailand linked their PayNow and PromptPay real-time retail payments system in April 2021, allowing customers to transfer funds simply by entering the recipient's phone number. Thailand then linked PromptPay to Malaysia's DuitNow fast-payments system, and Singapore and Malaysia announced plans to link PayNow and DuitNow in 2022. These real-time retail payments systems are organized by participating retail banks, organized through their respective national bankers' associations, with support from the central bank. Payments do not involve blockchain, though the Monetary Authority of Singapore and Bank Negra Malaysia have said they will consider integrating features such as distributed-ledger technology. In addition, credit card companies such as Visa and Mastercard, which operate in multiple countries, are developing the capability to settle transactions using stablecoins. In mid-2021 Mastercard announced a partnership with Circle, the principal issuer of USD Coin, which will enable it to accept USD Coin from card issuers and then either pay it out or the exchange it for fiat currency when settling with the merchant.

All this suggests that a variety of private entities are starting to do in the cross-border sphere the same things to which potential CBDC issuers aspire.

For whether CBDCs can be used outside the issuing jurisdiction, in the manner of the dollar today, central banks would have permit nonresidents to maintain digital wallets. In the PBOC's pilot projects to date, such permission has been promised only to foreigners temporarily traveling in China. Even if permission was granted, one wonders whether foreigners would feel safe using the Chinese CBDC, given privacy concerns. In mid-2021, the PBOC described "anonymous" wallets tagged only with a phone number (presumably a Chinese number), with balances limited to 10,000 yuan (US$1,560), but also wallets permitting larger balances and payments but requiring "valid ID" and bank account information (Phillips 2021). How comprehensively such transactions will be tracked by the authorities – how much information they will demand or harvest – is unclear. PBOC (2021) states that it will follow the principle of "anonymity for small value and traceable for high value." It insists that its CBDC "collects less transaction information than traditional electronic payment" and that the information so collected will not be shared with other central bank or government departments.

Alternatively, cross-border payments would be facilitated if different national CBDCs were interoperable. A growing number of central banks are investigating this possibility. For example, the Bank of Thailand and Hong Kong Monetary Authority are exploring building their own separate CBDC platforms ("Inthanon" and "LionRock") but allowing them to "talk to" one another.[25] Thus a Hong Kong importer of silk would be pay the Thai exporter in HK$, assuming that nonresidents are permitted to download a Hong Kong wallet. But that Thai exporter presumably has no appetite or use for HK$. An alternative would be for the Hong Kong importer to ask its bank for a HK$ depository receipt, at which point a corresponding amount of HK$ in the payer's account would be extinguished. That depository receipt would then be transferred into a dedicated international "corridor" where it would be exchanged for a Thai-denominated depository receipt at the best rate offered by dealers licensed to operate in the corridor. Finally the Thai payee's account would credited with the corresponding number of digital baht, extinguishing the depository receipt. The transaction would be completed in real time at a fraction of the current cost of cross-border payments.

---

[25] In a second project underway at the time of writing, the central banks of Hong Kong, Thailand, China and the United Arab Emirates, each with separate CBDC instructures, are exploring the possibility of making them interoperable.

Notice the preconditions for making this work. The two central banks would have to agree on an architecture for their digital corridor. They would have to jointly govern its operation. They would have to license and regulate dealers holding inventories of currencies and depository receipts to ensure that the exchange rate inside the corridor doesn't diverge significantly from that outside. They would have to agree on who provides additional liquidity, against what collateral, in the event of an order imbalance.

In a world of 200 currencies, moreover, arrangements of this type would require scores of bilateral agreements. And corridors of more than two countries would require rules and governance arrangements considerably more elaborate than those of the World Trade Organization or the IMF.

Finally, it is worth asking again: by how much would such arrangements reduce costs and increase speed relative to, say, SWIFT Go or blockchain-free linked instant payments systems a la Singapore and Thailand? With linked CBDC platforms, it would still be necessary to pre-validate or ex-post verify the identity of the customer account at the recipient bank. It would still be necessary to engage the services of an authorized dealer to complete the foreign exchange (depository receipt for depository receipt) transaction. One can imagine using automated market-making (AMM) and automated liquidity management (ALM) technology for the foreign exchange transaction, but these mechanisms have yet to be stress tested.[26] And it is not obvious why, if and when AMM and ALM technology is proven, it can't be adopted equally by SWIFT and other non-distributed-ledger-based services.

The alternative to linking separate national blockchains would be for multiple central banks to share a single blockchain. The Monetary Authority of Singapore and Banque de France have run experiments using Ethereum's permissioned enterprise blockchain. In the summer of 2021, the BIS announced that the MAS, Reserve Bank of Australia, Bank Negara Malaysia and South African Reserve Bank would engage in cross-border settlement trials using "a variety of different blockchain technologies and governance structures."

"A variety of different governance structures" leaves important questions up in the air. The type of governance structure that would be needed for a single unified blockchain running the currencies of 200 different countries kind of boggles the mind.

## 6. Conclusion

Digital currencies are coming, and nowhere faster than in Asia. The leading cyptocurrency exchange on which the stablecoin Tether is traded is based in Hong Kong. The Bank for International Settlements Innovation Hub, which coordinates official efforts in this area, is headquartered in Singapore. The first successful linkage of two national instant payments systems is between Singapore and Thailand.  The first experimental efforts to link separate blockchain-based central bank digital currencies is between Thailand and Hong Kong.  And China is set to be the first major country to issue a central bank digital currency.

---

[26] Automated liquidity management systems are programmed to provide rewards (additional digital tokens) for agents that lend the token in question when demand rises. As in old-fashioned systems of liquidity provision, one can imagine circumstances when there is no rate of return (no number of additional tokens) that compensate providers adequately for supplying such liquidity. Perhaps the central bank as liquidity provider of last resort can be programmed into such a system. Who knows?

Asia, clearly, is investing heavily in the digital currency sphere.  It is important to bear in mind, therefore, that significant uncertainties continue to dog digital currency initiatives.  Insofar as private-label stablecoins raise consumer protection, market integrity and systemic stability issues, will they be subject to oversight by the relevant financial supervisory authorities, be required to take out bank charters, or even be regulated out of existence?  Will stablecoins required to hold 100 percent collateral in the form of cash and short-duration Treasury securities and to impose redemption gates and fees ever amount to more than high-tech money market mutual funds?  Do the benefits of central bank digital currencies justify the costs?  Will CBDCs significantly reduce the cost and increase the speed of payments relative to non-blockchain-based instant payments and other digital innovations already pursued by other financial institutions?  And in terms of cross-border payments, can the obstacles to making CBDCs interoperable be successfully overcome?

We don't know the answers to these questions.  But we think that it is incumbent on Asian policy makers to address them.

# References

Attorney General of the State of New York, Investor Protection Bureau (2021), "Investigation by Letitia James, Attorney General of the State of New York, of iFinex Inc., BFXNA Inc., BFXWW Inc., Tether Holdings Limited, Tether Operations Limited, Tether Limited, Tether International Limited,"

Diem Association (2020), "Whitepaper V.2.0" (April) https://www.diem.com/en-us/white-paper/

Diamond, Douglas and Philip Dybvig (1983), "Deposit Insurance and Liquidity," *Journal of Political Economy* 91, pp.401-419.

Eichengreen, Barry (2021a), "Will Central Bank Digital Currencies Doom Dollar Dominance?" *Project Syndicate* (9 August).

Eichengreen, Barry (2021b), "The Stablecoin Illusion," *Project Syndicate* (13 July).

EU Blockchain Observatory and Forum (2021), "Central Bank Digital Currencies and a Euro for the Future," Brussels: European Commission.

Group of Thirty (2021), "U.S. Treasury Markets: Steps Toward Increased Liquidity," New York: Group of Thirty.

Hampl, Filip and Lucie Gyonyorova (2021), "Can Fiat-Backed Stablecoins be Considered Cash or Cash Equivalents under International Financial Reporting Standards Rules?" *Australian Accounting Review,* https://doi.org/10.1111/auar.12344.

Kozhan, Roman, and Ganesh Viswanath-Natraj. "Decentralized Stablecoins and Collateral Risk." *WBS Finance Group Research Paper Forthcoming* (2021).

Lyons, Richard K and Ganesh Viswanath-Natraj, "What Keeps Stablecoins Stable?," Technical Report, National Bureau of Economic Research 2020.

Olivra-Juarez, D. and E. Huerta-Manzanilla (2019), "Forecasting Bitcoin Pricing with Hybrid Models: A Review of the Literature," *International Journal of Advanced Engineering Research and Science* 6, pp.161-164.

People's Bank of China (2021), "Progress of Research & Development of E-CNY in China," Working Group on E-CNY Research and Development (July).

Phillips, Tom (2021), "PBOC: Chinese Consumers will be Able to Store their Digital Currency in both Physical and Digital Wallets," NFCW.com (16 June ), https://www.nfcw.com/whats-new-in-payments/pboc-chinese-consumers-will-be-able-to-store-their-digital-currency-in-both-physical-and-digital-wallets/

Routledge, Bryan and Ariel Zetlin-Jones (2021), "Currency Stability Using Blockchain Technology," *Journal of Economic Dynamics and Control* 130.

**Figure 1: USDT Spot and Futures Prices**

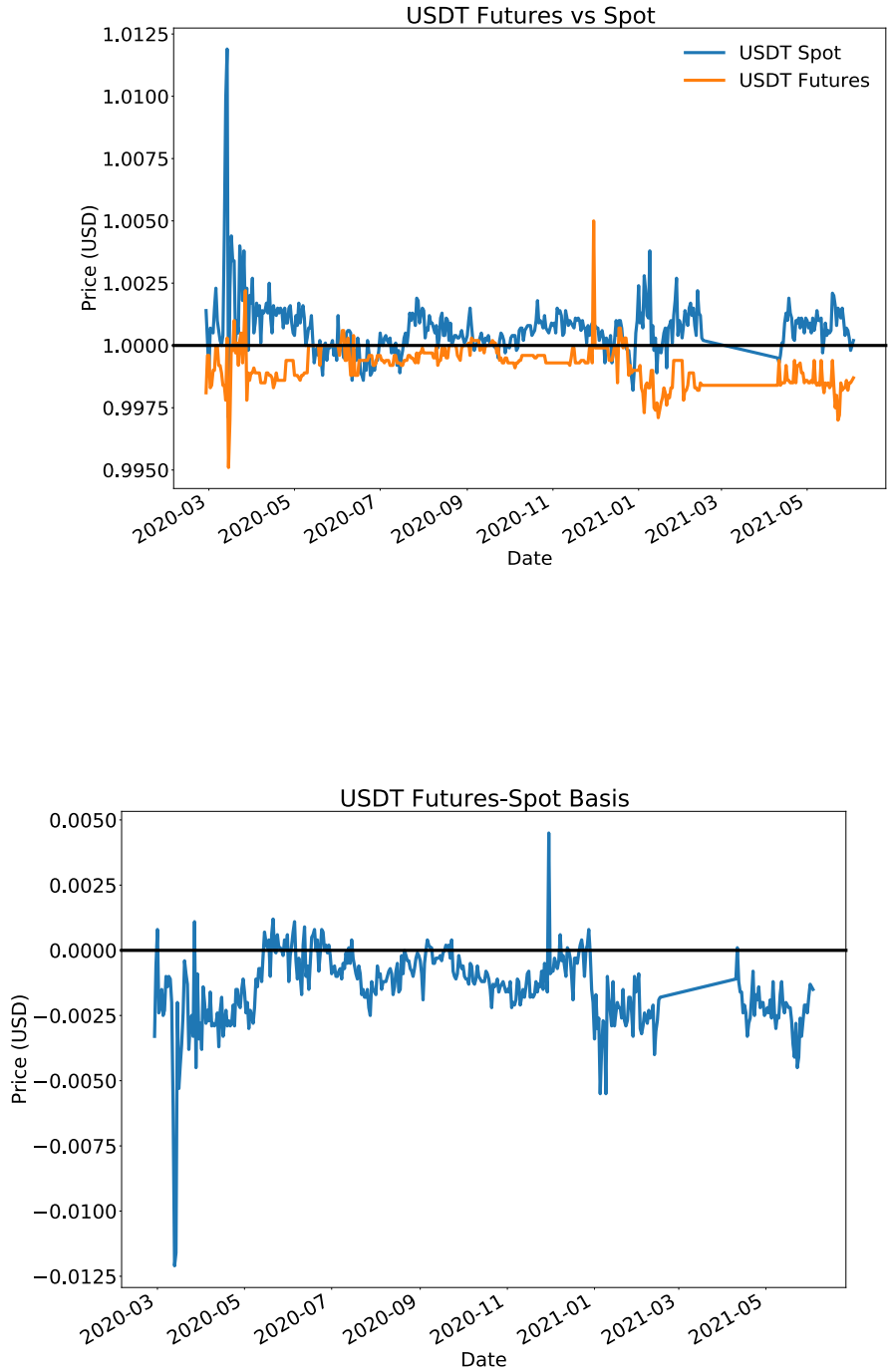

Note: Top panel: futures and spot prices on the FTX exchange. Bottom panel: Difference between futures and spot prices.

**Figure 2: Estimate of probability of default P**

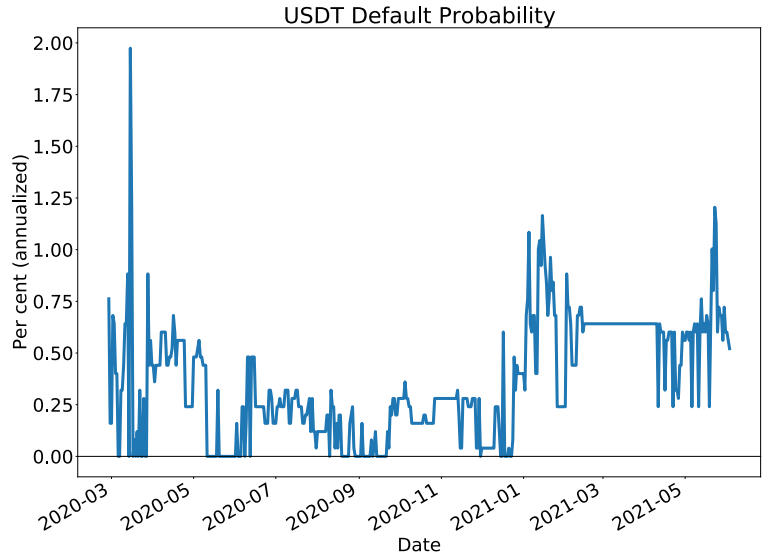

Note: Implied default probabilities based on spot, futures prices and the average mean reversion coefficient.

**Table 1: Stablecoins by Market Capitalization, October 2021**

| Stablecoin | Ticker | Market Cap |
|---|---|---|
| Tether | USDT | $70.32B |
| USDC | USDT | $32.98B |
| Binance USD | BUSD | $13.03B |
| DAI | DAI | $6.47B |
| Terra USD | UST | $2.78B |
| TrueUSD | TUSD | $1.18B |
| Paxos | PAX | $945.67M |
| Liquity USD | LUSD | $715.04M |
| Neutrino USD | USDN | $613.11M |
| HUSD | HUSD | $237.57M |

Note: Stablecoins by market cap, data source https://cryptoslate.com/cryptos/stablecoin/ accessed on October 30th, 2021.

**Table 2: Coinapi Data**

| Data Type | Coin Symbol | Exchange | Sample Period |
|-----------|-------------|----------|---------------|
| OHLCV | USDT_USD | FTX | 02/20-06/21 |
| OHLCV | USDT_USD | Kraken | 02/20-06/21 |
| OHLCV | BTC_USDT | Binance | 02/20-06/21 |

**Table 3: Summary statistics**

|  | count | mean | std | min | 25% | 50% | 75% | max |
|---|---|---|---|---|---|---|---|---|
| s | 410.0 | 1.0007 | 0.0011 | 0.9971 | 1.0001 | 1.0006 | 1.0011 | 1.0119 |
| f | 410.0 | 0.9992 | 0.0008 | 0.9951 | 0.9988 | 0.9993 | 0.9997 | 1.0050 |
| f-s | 410.0 | -14.2561 | 14.2087 | -121.0000 | -22.0000 | -13.0000 | -5.0000 | 45.0000 |
| P | 410.0 | 30.5791 | 31.4502 | -198.5050 | 12.0054 | 28.0294 | 48.0865 | 197.4453 |

**Table 4: Determinants of the Probability of Run Risk**

|  | I | II | III | IV |
|---|---|---|---|---|
|  | P | P | P | P |
| $\sigma_{BTC}$ | 0.0433*** |  |  | 0.0415*** |
|  | (0.0060) |  |  | (0.0064) |
| $\sigma_{USDT}$ |  | 0.2124*** |  | 0.0856 |
|  |  | (0.0702) |  | (0.0697) |
| $R_{BTC}$ |  |  | -0.0033 | 0.0008 |
|  |  |  | (0.0034) | (0.0033) |
| Intercept | 15.7474*** | 27.6135*** | 30.6252*** | 15.0208*** |
|  | (2.5363) | (1.8240) | (1.5617) | (2.6165) |
| R-squared | 0.11 | 0.02 | 0.00 | 0.12 |
| No. observations | 410 | 410 | 409 | 409 |

Note: This table regresses the probability of default P on intra-day volatility of BTC, USDT and BTC returns. The dependent variable in columns (I) through to (IV), P, measures the probability of default. The sample runs from February 28th, 2020 to June 1st, 2021. White heteroscedasticity-robust standard errors are in parentheses. *** denotes significance at the 1 percent level, ** at the 5 percent level, and * at the 10 percent level.